\documentclass[reqno,centertags, 12pt]{amsart}
\usepackage{amsmath,amsthm,amscd,amssymb,siunitx}
\usepackage{latexsym}
\usepackage{graphicx,wrapfig}
\usepackage{enumitem}
\usepackage{hyperref}
\usepackage[mathscr]{eucal}
\usepackage{tikz}
\usepackage{url}
\newcommand{\bbC}{{\mathbb{C}}}

\newcommand{\bbR}{{\mathbb{R}}}

\newcommand{\calH}{{\mathcal H}}

\newcommand{\bdone}{{\boldsymbol{1}}}

\newcommand{\lb}{\label}

\newcommand{\ran}{\text{\rm{ran}}}

\newcommand{\bi}{\bibitem}

\newcommand{\beq}{\begin{equation}}
\newcommand{\eeq}{\end{equation}}
\newcommand{\ba}{\begin{align}}
\newcommand{\ea}{\end{align}}

\newcounter{smalllist}

\newcommand{\comm}[1]{}

\DeclareMathOperator{\Real}{Re}

\allowdisplaybreaks
\numberwithin{equation}{section}

\newtheorem*{p2.1}{Proposition 2.1}

\theoremstyle{definition}

\newcommand{\jap}[1]{\langle #1 \rangle}
\newcommand{\norm}[1]{\lVert#1\rVert}

\AtBeginDocument{%
  \LetLtxMacro{\TheRealLabel}{\label}%
  \LetLtxMacro{\TheRealRef}{\ref}%
  \LetLtxMacro{\TheRealPageRef}{\pageref}%
}

\begin{document}

\title[Kato's Work]{Tosio Kato's Work on Non--Relativistic Quantum Mechanics: An Outline}
\author[B.~Simon]{Barry Simon}

\date{\today}
\maketitle

\centerline{IBM Professor of Mathematics and Theoretical Physics, Emeritus}

\centerline{California Institute of Technology, Pasadena, CA 91125}

\centerline{E-mail: bsimon@caltech.edu}

\section{Introduction} \lb{s1}

In 2017, we are celebrating the $100^{th}$ anniversary of the birth of Tosio Kato (August 25, 1917--October 2, 1999), the founding father of the theory of Schr\"{o}dinger operators.  There was a centennial held in Tokyo in his memory and honor in September.  I decided to write a review article on his work in nonrelativistic quantum mechanics (NRQM), which, as well see was only part of his opus.  I originally guessed it would be about 80 pages but it turned out to be more than 210!  It will appear in Bull. Math. Sci. \cite{FullKatoRev} with a shorter version also available \cite{ShortKatoRev}. I gave two lectures on Kato's work at the conference in his honor and this is an outline of what I said in those lectures.

The rest of this introduction will say a little about Kato's life while the next will summarize some major themes in his work.  I will then describe in some depth (but less detail than in my Bull. Math. Sci. article) five topics that were among the most important of Kato's contributions to NRQM.

Kato's most significant paper was on self--adjointness of atomic Hamiltonians published in 1951 in Trans. A.M.S. (see Section \ref{s3}).  I note that he was 34 when it was published (it was submitted a few years earlier, as we'll discuss in Section \ref{s3}). Before it, his most important work was his thesis, awarded in 1951 and published in 1949-51. One might be surprised at his age when this work was published but not if one understands the impact of the war. Kato got his BS from the University of Tokyo in 1941 but during the war, he was evacuated to the countryside. We were at a conference together one evening and Kato described rather harrowing experiences in the camp he was assigned to, especially an evacuation of the camp down a steep wet hill. He contracted TB in the camp. In his acceptance for the Wiener Prize, Kato says that his work on essential self{adjointness and on perturbation theory were essentially complete by the end of the war.

In 1946, Kato returned to the University of Tokyo as an Assistant (a position common for students progressing towards their degrees) in physics, was appointed Assistant Professor of Physics in 1951 and full professor in 1958.  Beginning in 1954, Kato started visiting the United States. This bland statement masks some drama. In 1954, Kato was invited to visit Berkeley for a year, I presume arranged by F. Wolf. Of course, Kato needed a visa and it is likely it would have been denied due to his history of TB. Fortunately, just at the time (and only for a period of about a year), the scientific attach\'{e} at the US embassy in Tokyo was Otto Laporte (1902-1971) on leave from a professorship in Physics at the University of Michigan. Charles Dolph (1919-1994), a mathematician at Michigan, learned of the problem and contacted Laporte who intervened to get Kato a visa. Dolph once told me that he thought his most important contribution to American mathematics was his helping to allow Kato to come to the US.

During the mid 1950s, Kato spent close to three years visiting US institutions, mainly Berkeley, but also the Courant Institute, American
University and Caltech. In 1962, he accepted a professorship in Mathematics from Berkeley where he spent the rest of his career and remained after his retirement. One should not underestimate the courage it takes for a 45 year old to move to a very different culture because of a scientific opportunity. The reader can consult the Mathematics Genealogy Project for a list of Kato's students (24 listed there, 3 from Tokyo and 21 from Berkeley; the best known are Ikebe and Kuroda from Tokyo and Balslev and Howland from Berkeley) and \cite{AMSMem} for a memorial article with lots of reminisces of Kato.

Kato received the Wiener prize in 1980 and passed away in 1999.

One gets some sense of the impact of Kato's work by looking at the number of things named after him which include:
\begin{enumerate}
  \item Kato's Theorem (on self--adjointness of atomic Hamiltonian)
  \item Kato--Rellich Theorem (stability of self adjointness under relatively bounded perturbations)
  \item Kato--Rosenblum Theorem and Kato--Birman theory (trace class scattering)
  \item Kato dynamics (adiabatic theorem)
  \item Kato cusp condition
  \item Trotter--Kato Theorem (3 different results!!)
  \item Putnam--Kato theorem (positive commutator $\Rightarrow$ pure ac spectrum)
  \item Kato smoothness theory
  \item Kato class
\end{enumerate}

There are several things called Kato's inequality

\begin{enumerate}
  \item His distributional inequality on $\Delta|u|$ (see Section \ref{s5})
  \item His inequality $\int |x|^{-1} |\varphi(x)|^2 \, d^3x \le \frac{\pi}{2} \int |k| |\hat{\varphi}(k)|^2 \, d^3k$
  \item A result on hyponormal operators that comes from Kato smoothness theory
  \item A variant of the Heinz--Loewner inequality on monotonicity of the square root that holds for maximal accretive operators.
\end{enumerate}

\section{Overview} \lb{s2}

One can roughly divide Kato's extensive research opus into four distinct areas:

\begin{enumerate}
  \item NRQM
  \item Nonlinear equations of mathematical physics including Euler, Navier--Stokes, KdV and non--linear Schr\"{o}dinger
  \item Theory of linear semigroups on Banach spaces
  \item Misc. Functional Analysis
\end{enumerate}

There is a kind of phase transition in his work around 1980.  Before then, he mainly worked on NRQM with occasional papers in the other three.  Afterwards, mainly in Nonlinear Equations.  Its almost as if when the theory of Schr\"{o}dinger operators became overcrowded, he switched to another area in which he could be a pioneer.

Kato's non--linear work is substantial.    He has over 13,500 MathSciNet citations  (so far as I know the most citations of any author is just under 22,000).   The top three are some edition or printing of his famous book \emph{Perturbation theory for linear operators} \cite{KatoBk}.  The next 9 are nonlinear (if you include his paper, \#11, with Br\'{e}zis on applying Kato's inequality to Schr\"{o}dinger operators with complex potentials  which I think is there because they have some lemmas widely used in current work on non--linear equations).

Interestingly enough, the most quoted in NRQM (at least under MathSciNet's data which only counts references in math papers and since about 2000) might surprise some of you.   It's his joint paper with Arne Jensen \cite{JenKato} on time--decay.  By the way, I suspect many of you will be even more surprised when I tell you later what is Kato's most quoted paper on Google Scholar!

Of course, one can't compartmentalize all his work.  I include his famous paper on the Trotter product formula for general self--adjoint contraction semigroups in his NRQM work although it fits even better in the linear semi--group area and he has a non--linear version with Masuda.  Anyhow, talking about his NRQM work will keep us busy enough so I won't say anything about the other work.

In my article, I divide Kato's work on NRQM into five subareas

\begin{enumerate}
  \item Eigenvalue Perturbation Theory
  \item Self--Adjointness
  \item Eigenvalue Results (only 2)
  \item Spectral and Scattering Theory
  \item Three Other Gems
\end{enumerate}

Kato has only two papers on (3) and five for (5).  For the others, there are many, many papers.  Below, we will focus on five sets of ideas, two in are (2) (Sections \ref{s3} and \ref{s5}), two are in (4) (Sections \ref{s6} and \ref{s7}) and one in (5) (Section \ref{s4}).

Interestingly enough, we will not discuss the NRQM paper with the most MathSciNet references (see above) and we'll barely mention his paper \cite{KatoEig} with the most Google Scholar references.  But I believe we'll discuss his most significant works.

\section{Foundations of Atomic Physics} \lb{s3}

Ever since the work of von Neumann about 1930, it has been clear that self--adjointness of quantum Hamiltonians is crucial.  In this regard, Kato has been a, indeed, the key figure.  His contributions include

\begin{enumerate}
  \item Self--adjointness of atomic Hamiltonians \cite{KatoHisThm}
  \item His work with Ikebe \cite{IK} on $V(x) > -cx^2-d$, an area where Weinholtz was the pioneer
  \item Kato's inequality, including his theorem that if $V \in L^2_{loc}(\bbR^\nu)$ and $V \ge 0$, then $-\Delta+V$ is essentially self--adjoint on $C_0^\infty(\bbR^\nu)$ (see Section \ref{s5})
  \item He was a pioneer on the use of quadratic forms including his work on perturbations (KLMN theorem) and monotone convergence for forms.
\end{enumerate}

In 1951 Kato published a theorem \cite{KatoHisThm} that he proved by 1944.   $N$--body quantum Hamiltonians on $L^2(\bbR^{3N})$ with two body potentials in $L^2(\bbR^3)+L^\infty(\bbR^3)$ are essentially self--adjoint on $C_0^\infty(\bbR^{3N})$ and self--adjoint on $D(-\Delta)$.    Since that class of potentials includes the Coulomb potential, this includes atomic and molecular Hamiltonians.

This is one of the most significant theorem in mathematical physics.   It makes Kato the father of the theory of Schr\"{o}dinger operators.   As Kato remarks in his Wiener prize acceptance, the proof is rather simple.   It relies on three steps.  First the Kato--Rellich theorem that if $A$ is self--adjoint and $B$ is $A$--bounded with relatively bound $a<1$, then $A+B$ is self--adjoint for $A$ and any core for $A$ is a core for $B$.

Secondly, by either a Sobolev estimate (or for Coulomb $V$, a Hardy inequality), any $L^2(\bbR^3)$ function, as an operator on $L^2(\bbR^3)$, is $-\Delta$--bounded with relative bound zero.   Finally, by integrating out the extra variables, relative boundedness on $\bbR^3$ implies relative boundedness of the pair potentials as operators on $\bbR^{3N}$.

It remains surprising that this theorem wasn't found earlier by Rellich or Friedrichs.   One factor is that von Neumann thought the problem was impossibly hard which may have discouraged people.   Rellich certainly knew the Kato--Rellich theorem by about 1940 and he used Hardy's inequality in one of his papers at that time.    All I can think of is that he missed the ``trivial'' idea of integrating out the extra variables  (which isn't trivial until you realize it!)

For our discussion of Kato's inequality below, we'll need one notion from later work extending Kato's Theorem.   Namely for general dimension $\nu$, the replacement of Kato's $L^2$ for $\bbR^3$ is $L^p$ where

$$ p = \left\{
         \begin{array}{ll}
           2, & \hbox{ if } \nu\le 3 \\
           >2, & \hbox{ if }\nu=4 \\
           \nu/2 & \hbox{ if } \nu \ge 5.
         \end{array}
       \right.$$

 For $V$'s of general sign, this is optimal since limit point/limit circle methods show if $\nu \ge 5$, then for $\lambda$ large, $-\Delta-\lambda |x|^{-2}$ is not self--adjoint on $D(-\Delta)$.   Note that $|x|^{-2} \in L^p+L^\infty$ for all $p < \nu/2$.

 The paper \cite{KatoHisThm} mentions that he can prove that eigenfunctions of atomic Hamiltonians are bounded.  He proved the details and much more in his paper \cite{KatoEig}.  Included was the Kato cusp condition which has been widely used in the atomic physics and quantum chemistry communities (which is why this has the most Google Scholar citations).

\section{The Adiabatic Theorem} \lb{s4}

In 1950, Kato published in J. Phys. Soc. Japan a paper \cite{KatoAdi} on the adiabatic theorem that has been central to the vast literature on the subject since then.  This subject concerns a family of time dependent Hamiltonians, $H(s), \, 0 \le s \le 1$ (which we'll assume is a set of bounded operators).   One wants to run the dynamics very slowly.  What that means is that one fixes $T$ large and looks at $H(s/T),\,0 \le s \le T$ and one wants to solve

\begin{equation*}
    \frac{d}{ds} \tilde{U}_T(s) = -iH(s/T)\tilde{U}_T(s),\,\, 0 \le s \le T; \qquad \tilde{U}_T(0) = \bdone
\end{equation*}

Letting $U_T(s) = \tilde{U}_T(sT), \, 0 \le s \le 1$, we see that $U_T(s),\,0\le s \le 1$ solves

\begin{equation*}
  \frac{d}{ds} U_T(s) = -iT H(s) U_T(s),\,\, 0 \le s \le 1; \qquad U_T(0) = \bdone
\end{equation*}

Kato supposed that $H(s)$ has an eigenvalue, $\lambda(s)$, which is continuous in $s$ and an isolated point of the spectrum of $H(s)$. While Kato was not explicit about technical hypotheses, it is easy to see that he needed to assume that $H(s)$ is $C^2$.  Let $P(s)$ be the projection onto the eigenspace $\ker(H(s)-\lambda(s))$.  Then Kato proved the \emph{Adiabatic Theorem}

\begin{equation*}
   \lim_{T \to \infty} (1-P(s))U_T(s)P(0) = 0
\end{equation*}

In other words, if $\psi \in \ran P(0)$, then as $T \to \infty$, $U_T(s)\psi$ gets closer to $\ran P(s)$.

In 1928, Born and Fock \cite{BF}  proved the above theorem when $H(s)$ has purely discrete spectrum and $\lambda(s)$ is a simple eigenvalue.   The point of Kato's version is not merely the greater generality but that the proof gives more information.

Recall that the theorem says that $\lim_{T \to \infty} (1-P(s))U_T(s)P(0) = 0$.   Kato discovered a dynamics in which this formula is exact.   He called it the adiabatic dynamics and I'll call it the Kato dynamics.   Namely if $W(s)$ solves

\begin{equation*}
   \frac{d}{ds}W(s) = iA(s)W(s), \, 0 \le s \le 1; \qquad W(0)=\bdone
\end{equation*}

\begin{equation*}
   iA(s) \equiv [P'(s),P(s)]
\end{equation*}

Then $W(s)$ obeys

\begin{equation*}
   W(s)P(0)W(s)^{-1} = P(s) \Rightarrow
\end{equation*}

\begin{equation*}
   (1-P(s))W(s)P(0) = (1-P(s))P(s)W(s) = 0
\end{equation*}

He proved the theorem by also proving that

\begin{equation*}
   \norm{e^{-iT\int_{0}^{s} \lambda(s) \, ds}  U_T(s)^*W(s)P(0)-P(0)} = \textrm{O}(1/T)
\end{equation*}

The point is that Kato not only proved that $U_T(s)\psi$ stayed inside $\ran P(s)$, he told you the exact vector it was.  And in this regard Kato left something on the table that took 30 years to be found and appreciated.

The dynamics, $U_T$, has two parts: the phase $e^{-iT\int_{0}^{s} \lambda(s) \, ds}$ and the Kato dynamics, $W(s)$.   In fancy language, the set of $k$ dimensional subspaces of our Hilbert space (or a nice subset) is a manifold.   For each point in the manifold, we have the corresponding subspace.  This defines a natural complex vector bundle (and, in case $k=1$, which is interesting, a complex line bundle).  The map $W(s)$ defines a notion of parallel transport, i.e. a connection.

The connection has a holonomy -- in the line bundle case, there is a finite phase associated to going in a closed path.   And this holonomy is a line integral which can be written, using Stokes' theorem, as a surface integral the curl of the connection, called the curvature.   The phase was found by Michael Berry \cite{Berry} in 1983 and is called Berry's phase and I interpreted it \cite{SimonBerry} as holonomy in the underlying connection.  As noted by Avron--Seiler--Simon \cite{ASSTKNN} one can use Kato's work to compute the connection in the underlying bundle.  This bundle and connection is also behind the work of Thouless et al (TKNN) \cite{TKNN} used in their work on the quantum Hall effect for which Thouless recently got a Nobel prize.   I want to emphasize that Kato's work is very much underlying the mathematics behind Berry phase.

\section{Kato's Inequality} \lb{s5}

This section will discuss a self--adjointness method that appeared in Kato \cite{KI1} based on a remarkable distributional inequality.  Its consequences is a subject to which Kato returned to often with at least seven additional papers \cite{KI2, KI3, KIComp1, KIComp2, KI4, KIComp3, KILp}.  It is also his work that most intersected my own -- I motivated his initial paper and it, in turn, motivated several of my later papers.

I want to set the stage by reminding you about Kato's 1951 paper and its aftermath.   For this, two--body suffices, so I consider $-\Delta+V(x)$ on $L^2(\bbR^\nu)$.  We had the notion of $p$ being $\nu$--canonical which I remind you, for $\nu \ge 5$ means $\nu/2$. By the late 1950's, it was known for esa--$\nu$ (i.e. essential self--adjointness on $C_0^\infty(\bbR^\nu)$) one needed some kind of uniformly local $L^p$ condition with $p$ being $\nu$--canonical.    The example $V(x) = -\lambda|x|^{-2}$ for which esa--$\nu$ can fail shows that $p=\nu/2$ is optimal.

When I entered the subject about 1970, there was an implicit belief that this condition applied to both the positive and negative parts.  In hindsight, this should not have been the case!    It was known there was a global asymmetry: one can't have behavior worse than O$(-|x|^2)$ at spatial infinity without losing esa  but if $V \ge 0$ and say, locally bounded, one has esa no matter what the growth at infinity.

But it was assumed that for local singularities, there was no difference.   In fact, the simplest example shows this to be wrong!   Consider the case $V(x) = |x|^{-\beta}$, where to have $C_0^\infty \subset D(V)$, one needs $\beta < \nu/2$.    Limit point/limit circle methods prove esa--$\nu$ for all such $\beta$  (but for $-|x|^{-\beta}$, esa--$\nu$ fails if $\beta > 2$).   So there was no good reason for the belief but, in fact, there was great surprise when I discovered the result I now turn to.

One of the advantages of working in multiple areas is that there can be technical cross--pollination.  In the fall of 1971, I was working on two--dimensional constructive quantum field theory.  One of the tools that had been developed to get self--adjointness of cut--off Hamiltonians concerned what are called hypercontractive semigroups.   In 1970, H\o egh--Krohn and I \cite{SHK} had abstracted results of Segal \cite{SegalHyperC}, invented the name hypercontractive and found an extension for positive potentials.  Using, in addition, ideas of Nelson \cite{NelsonHyperC1} and Konrady \cite{Konrady}, I was able to prove that \cite{SimonPosPot} ($d\mu(x) = e^{-|x|^2/2}\,dx$)

\begin{equation*}
  V \in L^2(\bbR^\nu,d\mu)\,\&\, V\ge 0 \Rightarrow -\Delta+V \textrm{ is esa--}\nu
\end{equation*}

So the positive part can have $L^2$ singularities.   While  $L^2(\bbR^\nu,d\mu)$ allows pretty wild growth at infinity, it was natural to conjecture, as I did, that if $V \ge 0$ one only needs $V \in L^2_{loc}(\bbR^\nu)$.

I wrote this up and sent out a preprint in early 1972 and within a month got a letter back from Kato, then almost 55, with a proof \cite{KI1} of my conjecture!   He used totally new ideas depending on a distributional inequality that if $u, \Delta u \in L^1_{loc}(\bbR^\nu)$ (allowed to be complex valued)  and $\mathrm{sgn}(u)(x)$ is defined to $0$ if $u(x)=0$ and otherwise $|u(x)|/u(x)$, then
\begin{equation*}
  \Delta |u| \ge \Real\left[\mathrm{sgn}(u) \Delta u\right]
\end{equation*}

The proof isn't hard; one proves it for $u$ smooth by taking derivatives of the definition $|u|_\epsilon^2=u^*u+\epsilon^2$ and taking $\epsilon\downarrow 0$.   Then one uses mollifiers to get it for general distributions.  If now $V \in L^2_{loc}$ and $V \ge 0$, one lets $H = -\Delta+V \restriction C_0^\infty \ge 0$ and considers $u \in D(H^*)$ with $H^*u = -u$.   This equation is a distributional equality so by Kato's inequality
\begin{equation*}
   \Delta |u| \ge (\mathrm{sgn}(u))(V+1)u = |u|(V+1) \ge |u|
\end{equation*}

Thus $(-\Delta+1)|u| \le 0$  so since $(-\Delta+1)^{-1}$ has a positive integral kernel, one sees that as a distribution $|u| \le 0$.  So, we have proven that $\ker(H^*+1) = \{0\}$, so by a theorem of von Neumann, $H$ is esa.    This really was remarkable.   The master of operator techniques  had suddenly pulled a distributional rabbit out of his hat.

Kato's inequality is quite striking, so it is interesting to see what it is really saying.  In this regard, I proved \cite{SimonKI1} the following result for arbitrary positive self--adjoint operators, $A$, on $L^2(M,d\mu)$.    The following are equivalent

\begin{enumerate}
  \item $e^{-tA}$ is positivity preserving:
\begin{equation*}
         \forall u \in L^2,\, u\ge 0, t \ge 0 \Rightarrow e^{-tA}u \ge 0
\end{equation*}

  \item (Beurling--Deny criterion) $u \in Q(A) \Rightarrow |u| \in Q(A)$ and
\begin{equation*}
        q_A(|u|) \le q_A(u)
\end{equation*}
  \item (Abstract Kato Inequality) $u \in D(A) \Rightarrow |u| \in Q(A)$ and for all $\varphi \in Q(A)$ with $\varphi \ge 0$, one has that
\begin{equation*}
        \jap{A^{1/2}\varphi,A^{1/2}|u|} \ge \Real\jap{\varphi,\mathrm{sgn}(u) Au}
\end{equation*}
\end{enumerate}

This shows that the Kato inequality methods are not unrelated to my original hypercontractive result  and led to a direct semigroup proof of the $L^2_{loc}$ result \cite{SimonMaxMin}.

Kato also proved a version of his inequality when there is a $C^1$ magnetic potential $\overrightarrow{a}$
\begin{equation*}
   \Delta |u| \ge \Real\left[\mathrm{sgn}(u)(\overrightarrow{\nabla}-i\overrightarrow{a})^2u\right]
\end{equation*}

This eventually led me (with a suggestion of Nelson) to  what is known as diamagnetic inequalities \cite{SimonKI1, SimonMaxMin}. If $H(a,V) = -(\nabla-ia)^2+V$, then
\begin{equation*}
   |e^{-tH(a,V)}\varphi| \le e^{-tH(a=0,V)}|\varphi|
\end{equation*}
pointwise.  There is a direct line from Kato's paper to this result.   As with the ordinary Kato inequality, there is an abstract semigroup version of this result found by Simon \cite{SimonKI2} and by Hess--Schrader--Unhlenbrock \cite{HessKI}.

\section{Kato--Rosenblum and Kato--Birman} \lb{s6}

There was an explosion of papers on time--dependent scattering theory in 1957--58 with work by Cook, Jauch, Kato and Rosenblum.  So far as I can tell, these were independent but several were clearly motivated by Friedrichs  \cite{FriedCont}.  In particular, Kato extended the notion of wave operator.   If $A$ and $B$ are arbitrary self--adjoint operators and $P_{ac}(X)$ the projection onto the absolutely continuous subspace for $X$, Kato \cite{KatoFR} considered

\begin{equation*}
   \Omega^{\pm}(A,B) = \textrm{s-}\lim_{t \to \mp \infty} e^{itA}e^{-itB}P_{ac}(B)
\end{equation*}
(the crazy $\pm$ vs. $\mp$ convention is from the physics literature and not what Kato used).  If these limits exist, one says the wave operators exist and if $\ran \, \Omega^{\pm}(A,B) = \ran P_{ac}(A)$, Kato called the wave operators complete.

Kato proved a basic fact:  $\Omega^{\pm}(A,B)$ are complete $\iff \Omega^{\pm}(B,A)$ exist.   The proof is almost trivial but this result is nevertheless very important since it reduces completeness to an existence property.   For example, motivated by this theorem, Deift and I (in 1977) showed \cite{DeiftSimonWO} that completeness of general $N$--body quantum scattering was equivalent to the existence of certain geometrically defined ``inverse'' wave operators.   So far as I know, every proof of $N$--body completeness uses the Deift--Simon wave operators.

One consequence of this theorem is that existence under a symmetric condition on $A$ and $B$ implies both existence and completeness.  In the same paper where the above fact appears, Kato proved that wave operators exist (and are complete) if $A-B$ is finite rank.

In short order after the finite rank results (in terms of paper publication dates), Rosenblum \cite{RosenTrace} proved that if $A$ and $B$ both have purely a.c. spectrum and $A-B$ is trace class, then the wave operators provide a unitary equivalence.  After that Kato \cite{KatoTC} proved what is often called the Kato--Rosenblum theorem:  if $A-B$ is trace class, then the Kato wave operators exist and are complete.

In terms of the sequence of papers, I'd always assumed that Rosenblum's paper was a rapid reaction to Kato's finite rank paper which, in turn, motivated Kato's trace class paper.   But I recently learned that is wrong.   Rosenblum was a graduate student of Wolf at Berkeley who submitted his thesis in March 1955.  It contained his trace class result with some additional technical hypotheses; a Dec. 1955 Berkeley technical report had his full result.  Rosenblum submitted a paper to the American Journal of Mathematics which took a long time refereeing it before rejecting it.   In April 1956, Rosenblum submitted a revised paper to the Pacific Journal in which it eventually appeared (this version dropped the technical condition; I've no idea what the original journal submission had).   So Rosenblum's work was independent of and (presumably) earlier than Kato's.

Kato's finite rank paper was submitted to J. Math. Soc. Japan on March 15, 1957 and was published in the issue dated April, 1957(!). The full trace class result was submitted to Proc. Japan Acad. on May 15, 1957.  Kato's first paper quotes an abstract of a talk Rosenblum gave to an A.M.S. meeting but I don't think that abstract contained many details.   This finite rank paper has a note added in proof thanking Rosenblum for sending the technical report to Kato, quoting its main result and saying that Kato had found the full trace class results (``Details will be published elsewhere.'').   That second paper used some technical ideas from Rosenblum's paper.

I've heard that Rosenblum always felt that he'd not received sufficient credit for his trace class paper.  There is some justice to this.    The realization that trace class is the natural class is significant.    In this regard, it is important that Weyl \cite{WeylvN} and von Neumann \cite{vNWeylvN} proved that for any self--adjoint, $A$, there is a Hilbert--Schmidt operator, $C$, so that $A+C$ has only point spectrum (so there is no chance that wave operators exist if $A$ has any a.c. spectrum).   And Kato's student, S. T. Kuroda, extended this \cite{KurodaWvN} to allow $C$ in any trace ideal other than trace class.

Kato was at Berkeley in 1954 when Rosenblum was a student (albeit some time before his thesis was completed)  and Kato was in contact with Wolf.   However, there is no indication that Kato knew anything about Rosenblum's work until shortly before he wrote up his finite rank paper when he became aware of Rosenblum's abstract.    My surmise is that both, motivated by Friedrichs, independently became interested in scattering.

I end by noting some of the important later developments (for references, see \cite{FullKatoRev}):

\begin{itemize}
  \item The original proofs were time--independent.  In his book Kato found a lovely time--dependent approach.
  \item Invariance principle and Two--Hilbert space scattering -- both of them where Kato made important contributions
  \item Extensions by Kuroda and especially Birman (so much so it is called Kato--Birman)
  \item Pearson's formulation which encompasses most extensions
  \item Krein spectral shift and the Birman--Krein formula
\end{itemize}

\section{Kato Smoothness} \lb{s7}

For the 25 years starting in 1955, spectral and scattering theory were one of Kato's major focuses. For example, when he was invited to give one of the plenary talks at the 1970 ICM, he chose \emph{Scattering theory and perturbation of continuous spectra} \cite{KatoICM} as his subject.

There are two approaches to proving results for scattering theory: time--dependent and time--independent  based, respectively, on wave operators and resolvent boundary values.   For many years, time independent was regarded as the more powerful but the Enss work changed that attitude to some extent.

In 1966 and 1968, Kato published two remarkable papers  \cite{KatoSm1, KatoSm2} (to me, the 1951 self--adjointness paper is Kato's most significant work, Kato's inequality his deepest  and the subject I'm about to discuss his most beautiful).   One of the things that is so beautiful is that there isn't just a relation between the two approach -- there is an equivalence  that depends on the observation that ($R(z) = (H-z)^{-1}$ for $z \in \bbC\setminus\bbR$)

\begin{equation*}
   \int_{0}^{\infty} e^{-\epsilon t}e^{i\lambda t}e^{-iHt}\varphi \, dt = -iR(\lambda+i\epsilon)\varphi
\end{equation*}
this plus the Planceherel formula proves that
\begin{equation*}
   \int_{-\infty}^{\infty} \left(\norm{AR(\lambda+i\epsilon)\varphi}^2+\norm{AR(\lambda-i\epsilon)\varphi}^2\right) \,d\lambda =
\end{equation*}
\begin{equation*}
   2\pi \int_{-\infty}^{\infty} e^{-2\epsilon |t|} \norm{Ae^{-itH}\varphi}^2 \, dt
\end{equation*}

 We say that $A$ is $H$--smooth if there is $C < \infty$ so that for all $\epsilon>0$,  the left side of this is bounded by $(2\pi C)^2\norm{\varphi}^2$.

Thus we have the first two of a remarkable set of equivalences \cite{KatoSm1}:

\begin{enumerate}
  \item $A$ is $H$--smooth with constant $C$
  \item $\sup_{\norm{\varphi}=1} \int_{-\infty}^{\infty} \norm{Ae^{-itH}\varphi}^2 \, dt \le C^2$
  \item $\sup_{\norm{\varphi}=1, -\infty<a<b<\infty} \frac{\norm{AP_{(a,b)}(H)\varphi}^2}{b-a} \le C^2$
  \item $\sup_{\substack{\norm{\varphi}=1,\\ \mu \notin\bbR, \varphi \in D(A^*)}} (2\pi)^{-1} |2\,\textrm{Im} \jap{A^*\varphi,R(\mu)A^*\varphi}| \le C^2$
  \item $\sup_{\norm{\varphi}=1, \mu \notin\bbR, \varphi \in D(A^*)} (\pi)^{-1} \norm{R(\mu)A^*\varphi}^2|\textrm{Im}\mu| \le C^2$

\end{enumerate}

From (4) and Stone's formula, we immediately see that $\ran(A^*) \subset \calH_{ac}(H)$  so the existence of $H$--smooth operators has spectral consequence for $H$.

The uniformity in the time--dependent condition immediately implies that if $H=H_0+A^*B$  and if $B$ is $H_0$--smooth and $A$ is $H$--smooth,  then by noting the derivative is integrable, we see that $\Omega(H,H_0)$ exists.   Since $H_0=H-B^*A$, we also have the inverse wave operators exist and so completeness.

Kato--Yajima \cite{KY} (1989) say that $A$  is $H$--supersmooth if
\begin{equation*}
   \sup_{\substack{\norm{\varphi}=1,\\ \mu \notin\bbR, \varphi \in D(A^*)}} (2\pi)^{-1} |\jap{A^*\varphi,R(\mu)A^*\varphi}|<\infty
\end{equation*}
The notion (which appeared already in Kato \cite{KatoSm1}) remains useful but, alas, the name didn't stick.   In any event, the above implies  that the imaginary part is bounded so supersmoothness$\Rightarrow$smoothness.   More importantly, Kato \cite{KatoSm1} noted (as follows from a geometric series), that if the above sup is less than $1$ and if $\norm{C}\le 1$, then $A$ is also $H_1$--smooth if $H_1=H+A^*CA$  so the wave operators exist and are unitary.

Moreover, it is known (Kato--Yajima \cite{KY}, implicit in Kenig--Ruiz--Sogge \cite{KRS}) that if $V \in L^{\nu/2}(\bbR^\nu),\,\nu \ge 3$, then $|V|^{1/2}$ is $H$--supersmooth.   So for such $V$'s, one has a small coupling result (Kato's original paper had a slightly weaker result).   Iorio--O'Carroll \cite{IorioOC} (Iorio was Kato's student) have an analogous $N$--body weak coupling result.

In 1968, Kato \cite{KatoSm2} found that smooth perturbations are an ideal tool for proving a theorem found the year before by Putnam \cite{PutnPK} and now called the Kato--Putnam theorem:  if $A$ and $B$ are bounded self--adjoint operators so that $i[A,B]$ is strictly positive (i.e. for all $\varphi \ne 0$, we have that $\jap{\varphi,i[A,B]\varphi} > 0$),  then $A$ and $B$ both have purely a.c. spectrum.

For let $C$ be the square root of $i[A,B]$.  Then $\frac{d}{dt}\jap{e^{-itA}\varphi,Be^{-itA}\varphi}^2 = \norm{Ce^{-itA}\varphi}^2$ so the integral of $\norm{Ce^{-itA}\varphi}^2$ from $s$ to $t$ is bounded by $2\norm{B}\norm{\varphi}^2$.  Thus $C$ is $A$--smooth and $A$ has only a.c. spectrum on the closure of $\ran(C)$ which is all of $\calH$.

Further developments, discussed in \cite{FullKatoRev}, include Lavine's local smoothness theory and its application to repulsive potentials and a wonderful and not well-known application by Vakulenko.


\end{document}